# The benefits and challenges of explicit memory management in OpenMP Target GPU offloading

Igor Sfiligoi

Igor Sfiligoi, University of California San Diego, isfiligoi@sdsc.edu

OpenMP Target Offload is a popular GPU technology for porting compute codes that operate on large buffers. One of the main usability features that is typically emphasized is the semi-automatic handling of buffer synchronization in partitioned memory setups, typical of discreet GPU systems. That feature however comes with potential correctness, performance and resource consumption drawbacks. This paper outlines the drawbacks of that approach and outlines how explicit handling of buffer locality, natively supported through OpenMP, avoids most of those pitfalls, but also comes with its own downsides.

## 1 OPENMP TARGET IMPLICING MEMORY MANAGEMENT

In recent years most performance increases in large scale computing have come from GPU technologies, prompting scientific software developers to write codes that can execute on these compute resources. GPU computing does come with severe programming limitations in the general case, but it maps easily when solving data-parallel problems, which are typically addressed by looping over large data buffers in languages like C++ and Fortran. Note that this is a somewhat similar transition as the scientific compute community encountered when CPUs moved from being purely sequential resources to being massively multi-threaded. It is thus perhaps not a surprise that OpenMP [1], originally developed for CPU-based parallelization, can now address the parallelization needs of GPUs as well.

From a purely computational point of view, many OpenMP-based codes can be ported to GPU compute with a simple (potentially conditional) tweaking of the OpenMP pragmas, i.e. by telling the OpenMP runtime to execute the code portion on the GPU instead of doing it on the CPU, as shown in the Fortran-based Algorithm 1. Note that explicit offloading is needed, since OpenMP allows parallel-but-CPU-only and GPU-parallel code to co-exist in the same process.

ALGORITHM 1: Simple OpenMP Target Fortran subroutine

```
subroutine add_inc(n,A,B)
  integer, intent(in) :: n
  real, dimension(n), intent(inout) :: A,B
  integer :: i
!$omp target teams distribute parallel do
  do  i=1,n
     A(i) = A(i)+B(i)
     B(i) = B(i) + 1
  enddo
end subroutine
```

This basic approach does however come with a hefty performance penalty, since the OpenMP compiler will implicitly generate the necessary code to make a copy of the CPU-resident buffer into a dedicated GPU-resident buffer(s) before starting the GPU compute, and also the necessary code to copy the content of the buffer(s) back, after the GPU compute completes. This is typically needed, due to the partitioned memory nature of most discreet GPU-based systems.

To avoid this penalty, the OpenMP Target standard provides pragmas for managing a long-term copy of the CPU-resident buffers in the GPU memory partition. Moreover, the creation and/or synchronization of the two buffers can happen anywhere in the code and can even be done in objects compiled independently. While this addresses the performance concerns, if used correctly, it increases both the risk of the two buffers diverging, with some code sections using stale data and thus getting wrong results, and the risk of some GPU code sections inadvertently using buffers that do not have long-term GPU residency, resulting in hidden slowdowns.

Moreover, automatic buffer management, both pointer translation and optional buffer synchronization, only works for simple arrays. When either using non-trivial pointer arithmetic or accessing pointers in nested structures, as it is common practice in object-oriented C++, most compilers will not automatically handle the GPU translation correctly, likely resulting in illegal memory locations being accessed. To support such code sections, a programmer must provide comprehensive description of the object's internals to OpenMP as shown in Algorithm 2, thus completely circumventing the object-oriented abstraction layer.

ALGORITHM 2: C++ function processing an object with OpenMP Target data handling

```
void add_inc(int n,TVect& A, TVect& B) {
#pragma omp target teams distribute parallel for map(to:A,B) map(tofrom:A.buf[:n],B.buf[:n])
  for (int  i=0; i<n; i++) {
     A(i) = A(i)+B(i);
     B(i) = B(i) + 1;
  }
}
```

And, finally, there is also the issue of excessive resource consumption, namely the often unused CPU buffer. Given the superior performance of GPU compute, it is quite normal for only the GPU buffer to be in active use. Nevertheless, the CPU-resident buffer still must be allocated and alive for the full duration of the compute process, since the OpenMP Target runtime depends on it. This used to be a minor annoyance when GPU memory subsystems were a fraction of the memory available to the CPUs, but modern systems often have a comparable amount of memory in both the CPU and GPU subsystems, potentially resulting in problem sizes being artificially limited by CPU memory. This is further exacerbated in systems that actually share the memory pool between the CPU and GPU cores, e.g. the AMD MI300A APU [2], where buffer mirroring effectively cuts the available memory in half.

## 2 EXPLICIT GPU MEMORY ALLOCATION USING OPENMP

Buffer mirroring is however not a hard requirement, and the OpenMP Target standard does support GPU-only memory buffers. Of course, such buffers cannot be accessed from CPU-only code. But when a programmer expects only GPU code to access that data, it avoids both the waste of CPU memory and the risk of buffer content divergence.

The major downside of the GPU-only approach is programming complexity. In particular, users cannot use the language-standard buffer allocation methods, and must replace those calls with functions provided by OpenMP, which is significantly more complex. While this is arguably just a modest annoyance in C++, since it can be abstracted away using wrappers, it is much more intrusive in Fortran, where OpenMP-allocated buffers must be explicitly handled as pointers, as shown in Algorithm 3.

ALGORITHM 3: Example GPU-only allocation in Fortran

```
real, dimension(:,:,:,:), pointer :: buf
...
subroutine gpualloc(d1,d2,d2,d4)
  integer, intent(in) :: d1,d2,d3,d4
  type(c_ptr) :: c_ptr_buf
  integer(c_size_t) :: bsz
  bsz= d1*d2*d3*d4*4 ! real is fp32 == 4 bytes
  c_ptr_buf = omp_target_alloc(bsz, omp_get_default_device())
  call c_f_pointer(c_ptr_buf, buf, [d1, d2, d3, d4])
end subroutine
```

Moreover, the OpenMP runtime does not automatically recognize GPU-only buffers, and the user must explicitly label them as such in the OpenMP data pragmas. This can be potentially confusing in nested structures in C++, where the rule only applies to the top-level object, and not to any of the inner buffer pointers. Nevertheless, if used consistently, GPU-only buffers end up being much easier to use, especially when it comes to nested data structures in C++. Since those nested pointers are already pointing to GPU buffers, there is no need for the caller to know what's behind the abstraction layer, as shown in Algorithm 4, simplifying code development while preserving both correctness and performance, as there is never any implicit memory copy involved.

ALGORITHM 4: C++ function processing GPU-resident objects with OpenMP Target

```
void vect_add_inc(int n, int m, TVect* A, TVect* B) {
#pragma omp target teams distribute parallel for collapse(2) is_device_ptr(A,B)
  for (int  v=0; v<m; v++) for (int  i=0; i<n; i++) {
     A[v](i) = A[v](i)+B[v](i);
     B[v](i) = B[v](i) + 1;
  }
}
```

That said, mixing GPU-only and mirrored buffers is allowed, and sometimes needed, e.g. when some (summary) data has to be returned back to CPU-only memory, or when propagating CPU-only data into the GPU buffers during initialization. Those situations are, however, typically rare, so the tradeoff is acceptable.

## 3 SHARING ADDRESS SPACE BETWEEN CPU AND GPU CORES

Another option provided by the OpenMP Target standard is sharing of the address space between CPU and GPU cores. When enabled, OpenMP disables both pointer translation and buffer mirroring, resulting in both CPU-allocated and GPU-allocated buffers to be treated in the same way, effectively ignoring all of the data-related pragmas.

This paradigm is indeed the most user-friendly and least error prone, when it can be used. Unfortunately, the software itself cannot just magically stitch together independent memory subsystems, so support for the unified memory address space must be provided at the hardware and driver level. Most modern systems do have good support for this paradigm, but it is often not turned on by default, which may limit the range of targetable resources.

Moreover, many GPU-enabled libraries rely on the partitioned memory address space as an implicit parameter to switch between CPU and GPU compute. If the address space becomes unified, the semantics of those libraries is often not well defined. While arguably this should be fixed, it may result in unexpected problems on current systems.

## 4 SUMMARY AND CONCLUSIONS

Given that most GPU-enabled systems come with partitioned memory subsystems, the OpenMP Target default mirroring of CPU and GPU memory buffers is a great usability feature when dealing with simple buffers. But it can also lead to both correctness and performance issues if not used with great care, potentially leading to obscure problems in complex codebases, especially as that codebase evolves with time.

There has been a recent push for providing a unified memory view of the two subsystems, with a combination of hardware and driver capabilities, which would solve that issue at its root. Unfortunately, many of the currently deployed systems don't support that out of the box, limiting its potential appeal.

The alternative is to explicitly handle CPU-resident and GPU-resident buffers as separate entities. While this does result in some less-user-friendly code sections, especially during memory allocation and cleanup, it avoids a large class of potential correctness and performance issues. This is especially noticeable when nested data structures are used, which is typical in C++ programs, actually resulting in more user-friendly code patterns in the majority of the codebase.

One additional benefit of the explicit GPU buffer handling is the drastic reduction in CPU-associated memory use. This is important for both discreet GPU systems with modest system memory, and true shared memory APU systems.

## ACKNOWLEDGMENTS

This work was partially funded by the U.S. National Science Foundation, Office of Advanced Cyberinfrastructure grant OAC-2404323, and the U.S. Department of Energy, Office of Science, Office of Fusion Energy Sciences award DE-SC0024425.

## REFERENCES

[1] OpenMP Home Page, https://www.openmp.org, Accessed Mar 17th, 2026.

[2] AMD Instinct MI300A Accelerators, https://www.amd.com/en/products/accelerators/instinct/mi300/mi300a.html. Accessed Mar 17th, 2026.